\documentclass[]{spie}  
\usepackage[]{graphicx}

\title{Generic Misalignment Aberration Patterns and the Subspace of
  Benign Misalignment}

\author{Paul L. Schechter\supit{a,b} and Rebecca Sobel Levinson\supit{a,b}\footnote{~Rebecca Sobel Levinson is an NSF Graduate Research Fellow}
\skiplinehalf
\supit{a}MIT Kavli Institute, 77 Massachusetts Avenue, Cambridge, Massachusetts, USA \\
\supit{b}Department of Physics, Massachusetts Institute of Technology
}

\authorinfo{E-mail: schech@mit.edu \& rsobel@mit.edu}

\begin{document} 
  \maketitle 

\begin{abstract}
Q1: Why deploy $N$ wavefront sensors on a three mirror anastigmat (TMA)
and not $N + 1$?\\
Q2: Why measure $M$ Zernike coefficients and not $M + 1$?\\
Q3: Why control $L$ rigid body degrees of freedom (total) on 
    the secondary  and tertiary and not $L + 1$?\\
The usual answer:  ``We did a lot of ray tracing and $N, M,$ and $L$ seemed OK.''\\  
We show how straightforward results from aberration theory
may be used to address
these questions.  We consider, in particular, the case of a three mirror
anastigmat.
\end{abstract}

\keywords{Telescopes: wide-field, alignment, wavefront sensing}

\section{INTRODUCTION}
\label{sec:intro}  

\subsection{New Wide Field Telescopes and Wavefront Sensing}

The next decade will see the construction of several
billion-dollar-class wide field telescopes, ground-based and in space,
for which the measurement of cosmological weak gravitational lensing
is a major programmatic goal.\footnote{LSST, Euclid, WFIRST}  The
image quality delivered by these telescopes must be understood to a
level never before achieved by astronomical telescopes.

At least four different effects will contribute to the observed point
spread function (henceforth the PSF): telescope tracking errors, optical
misalignments, mirror figure errors, 
and diffusive and transfer effects within the detector.
For ground-based telescopes, atmospheric seeing makes an additional
contribution.

Wavefront sensors may be deployed to disentangle these effects, and in
particular, to measure mirror figure errors and the misalignments of
the telescope optics.  A recent study of the design for the Large
Synoptic Survey Telescope (henceforth the LSST) describes four
wavefront sensors measuring Zernike polynomials up to 36th
order\cite{Manuel2010}.  But why four, and why 36th order?  The answer
would appear to be that ray tracing was carried out for many
combinations, and that these choices sufficed.

While ray tracing simulations may identify wavefront sensing
configurations that are effective and efficient, one might still want
to understand {\it why} they are effective and
efficient.  And insofar as ray tracing simulations are only
approximate in their modeling of telescopes and detectors, an
understanding of what lies behind the answers can be quite useful.
The purpose of this paper is to present a coherent and concise
development of such an understanding.

\subsection{Outline of Paper}  

In \S2 we describe the generic  3rd order
Seidel misalignment aberration patterns that arise in circularly symmetric
telescopes.  In \S3 we consider the particular case of a
three mirror anastigmat, and show that the 3rd order misalignment 
aberration patterns do not suffice 
to produce perfect telescope alignment.
In \S4 we describe the generic  5th order
Seidel misalignment aberration patterns that arise in circularly symmetric
telescopes.
In \S5 we introduce the {\it subspace of benign
  misalignment} defined by the subset of misalignments that produce
none of the third order asymmetric aberration patterns described in \S
3.  In \S6 we answer, in part, the questions of
the number of wavefront sensors needed and the order to which the wavefront
must be measured.

\section{GENERIC MISALIGNMENT ABERRATION PATTERNS: 3RD ORDER}

The development outlined here draws heavily on that of 
Tessieres\cite{Tessieres2003}, which in turn draws heavily on that of 
Thompson and collaborators \cite{Thompson1980, Thompson2005, Thompson2009}.  
It is described in considerably more detail by
Schechter and Levinson\cite{Schechter2011}.

\subsection{Symmetric Third Order Aberration Patterns}

A parabolic one mirror telescope produces four different symmetric
third order Seidel aberration patterns at its prime focus: coma,
astigmatism, curvature of field and distortion.  A Ritchey-Chretien
two-mirror telescope is designed to eliminate the symmetric coma pattern at it
the focus of its secondary.  A three mirror anastigmat is designed to
remove both the symmetric coma and symmetric
astigmatism patterns at the focus of its
tertiary.\footnote{We shall ignore distortion patterns since these
produce only wavefront tilts, displacing the positions of otherwise perfectly
focused images.  One can imagine using preexisting astrometry and science
images to measure distortion patterns, which might obviate the
need for measuring one or more of the patterns described below.}

Though one may have designed a telescope to eliminate these patterns,
they reappear if one has despace errors.  Alternatively, if the design
hasn't eliminated one of these patterns, despace errors will change
its amplitude.

   \begin{figure}
   \begin{center}
   \begin{tabular}{c}
   \includegraphics[scale=0.52]{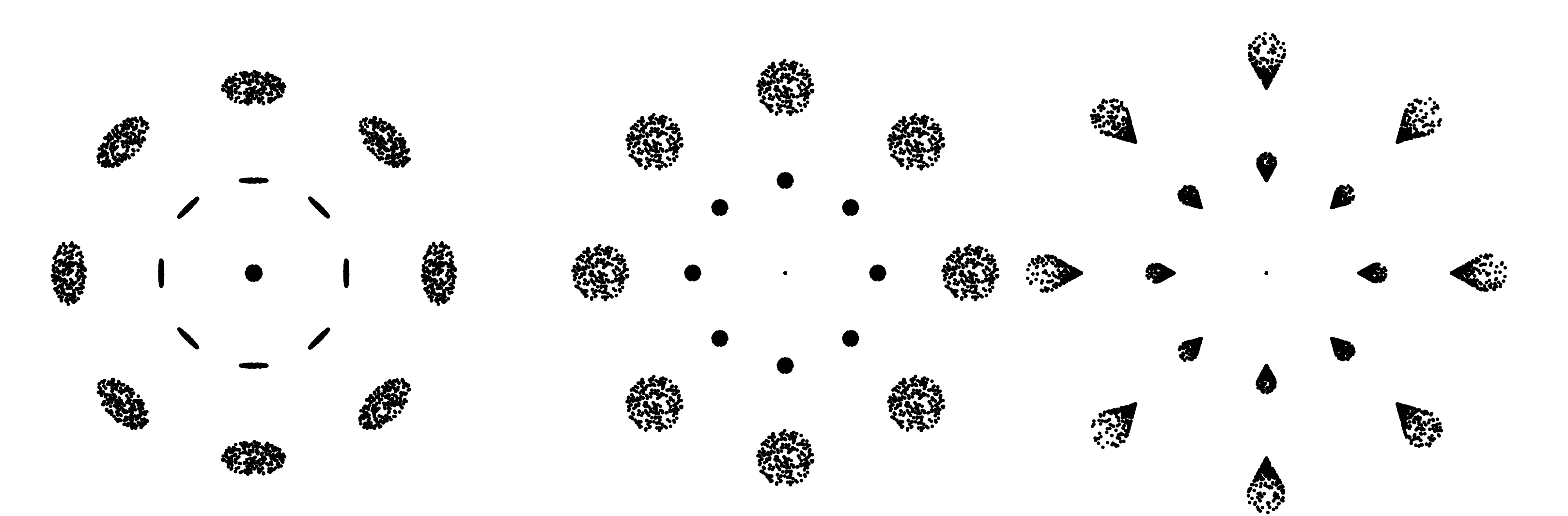}
   \end{tabular}
   \end{center}
   \caption[3sym] 
   {\label{fig:3sym} 
Symmetric 3rd order aberration patterns: astigmatism, curvature of
field and coma.  Spherical aberration (not shown) is constant across
the field.  A constant defocus has been added to the astigmatism to
show the orientations of the astigmatic images.
}
   \end{figure} 

In figure \ref{fig:3sym} we show the three symmetric third order Seidel aberration
patterns that produce imperfect images.  Note that some defocus has
been added to the astigmatism to show the orientation of the pattern.
Each symmetric pattern is described by one number, an amplitude.

\subsection{Asymmetric Third Order Aberration Patterns}

\begin{figure}
   \begin{center}
   \begin{tabular}{c}
   \includegraphics[scale=0.52]{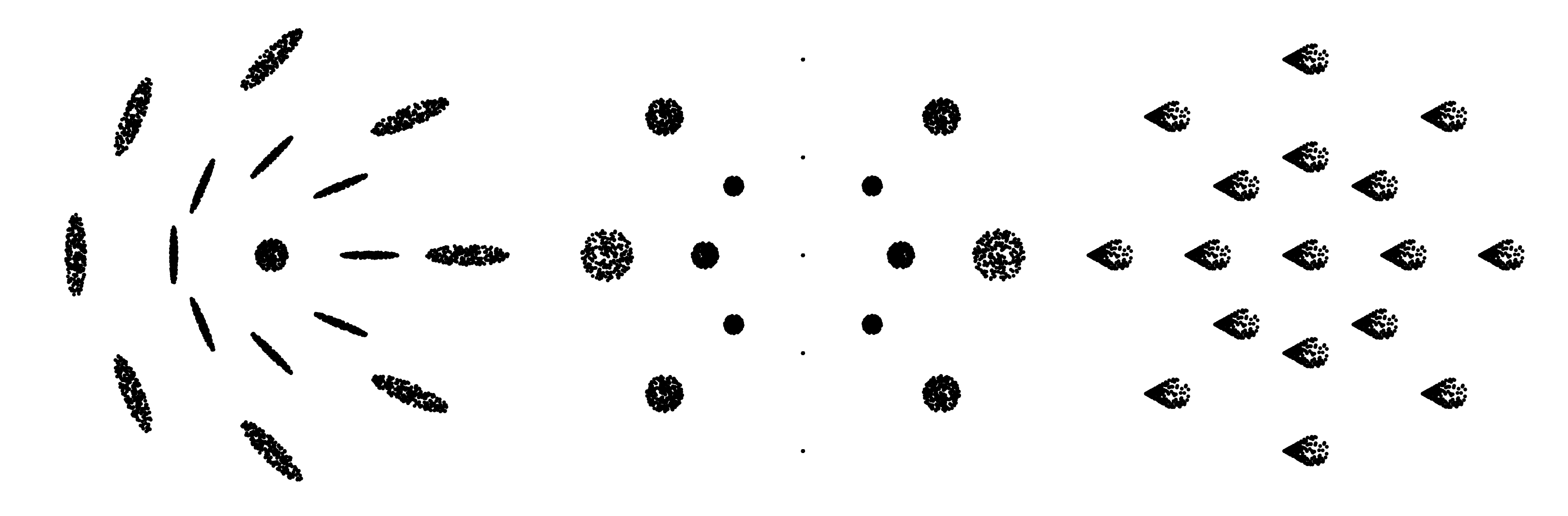}
   \end{tabular}
   \end{center}
   \caption[a3sym] 
   {\label{fig:3asym} 
Asymmetric 3rd order aberration patterns: astigmatism, curvature of
field and coma.  
}
   \end{figure}

Decenter and tilt errors produce asymmetric third order Seidel
aberration patterns.  In contrast with the symmetric patterns, each
asymmetric pattern is described by {\it two} numbers, an amplitude and
a direction.  These patterns are shown in figure \ref{fig:3asym}.  In general a misaligned
mirror will produce a superposition of these patterns.

\section{THE CASE OF THE THREE MIRROR ANASTIGMAT}

\subsection{Degrees of Freedom}

We consider the particular case of a circularly symmetric three mirror
anastigmat (TMA).  We take the position and orientation of the primary
to define the telescope pointing.  Assuming circular symmetry, there
are five rigid body degrees of freedom each for the secondary and
tertiary, for {\it a total of ten.}  One must therefore measure ten
numbers to keep such a telescope perfectly aligned.

\subsection{Corrected to 3rd Order but Still Imperfectly Aligned}

The two symmetric degrees of freedom that must be controlled are the piston
of the secondary and the tertiary.  There are five aberration
patterns that might be used to do this: third order coma, astigmatism,
curvature of field and spherical aberration and first order defocus.

This leaves eight asymmetric degrees of freedom.  Measuring asymmetric
coma, astigmatism and curvature of field (which manifests itself as a
tilt of the focal plane) leaves us with two uncorrected asymmetric
degrees of freedom.  The telescope is still imperfectly aligned.

While one might in principle measure the third order asymmetric
distortion patterns, these require a precision astrometric catalog.
The alternative is to measure one of the asymmetric fifth order
aberration patterns.

\section{GENERIC MISALIGNMENT ABERRATION PATTERNS: 5TH ORDER}

A three mirror anastigmat eliminates symmetric aberration patterns
only to third order.  There are two broad classes of symmetric fifth
order aberration patterns: coma, astigmatism, defocus and spherical
aberration that vary as a higher power of field angle and those that
vary more rapidly on the pupil.

\subsection{Nomenclature}

There is no consistent nomenclature for fifth order Seidel aberrations
We shall refer to the pattern of comatic images that varies as the third
power of field angle as {\it fifth order coma.}  Likewise we shall refer to
the patterns of astigmatic images, defocussed images, and spherically
aberrated images that vary by two powers of field angle more than
their third order counterparts as {\it fifth order astigmatism, fifth
order defocus} and  {\it fifth order spherical aberration.}  These
aberrations are all described by low order Zernike polynomials.

In addition to these, there are fifth order aberrations that have no
third order counterpart -- they produce spot patterns unlike those
of third order and are described by higher order Zernike polynomials.
An example would be the pattern that varies as $\rho^5 \cos\theta$ on
the pupil.  We shall refer to this as {\it coma-II} to distinguish it
from conventional coma (coma-I) that varies $\rho^3 \cos\theta$ on
the pupil.  Likewise we shall refer to the aberration that varies
as $\rho^4 \cos2\theta$ as {\it astigmatism-II}, and to the aberration
that varies as $\rho^6$ as {\it spherical-II}.\footnote{Were we 
completely consistent we would use the term
{\it focus-II} in preference to spherical aberration, but
this would create rather than eliminate confusion.}

Rounding out the complement of fifth order Seidel aberrations is
{\it trefoil}, which varies as  $\rho^3 \cos3\theta$ on the pupil.

\subsection{Symmetric Fifth Order Aberration Patterns}

\begin{figure}
   \begin{center}
   \begin{tabular}{c}
   \includegraphics[scale=0.52]{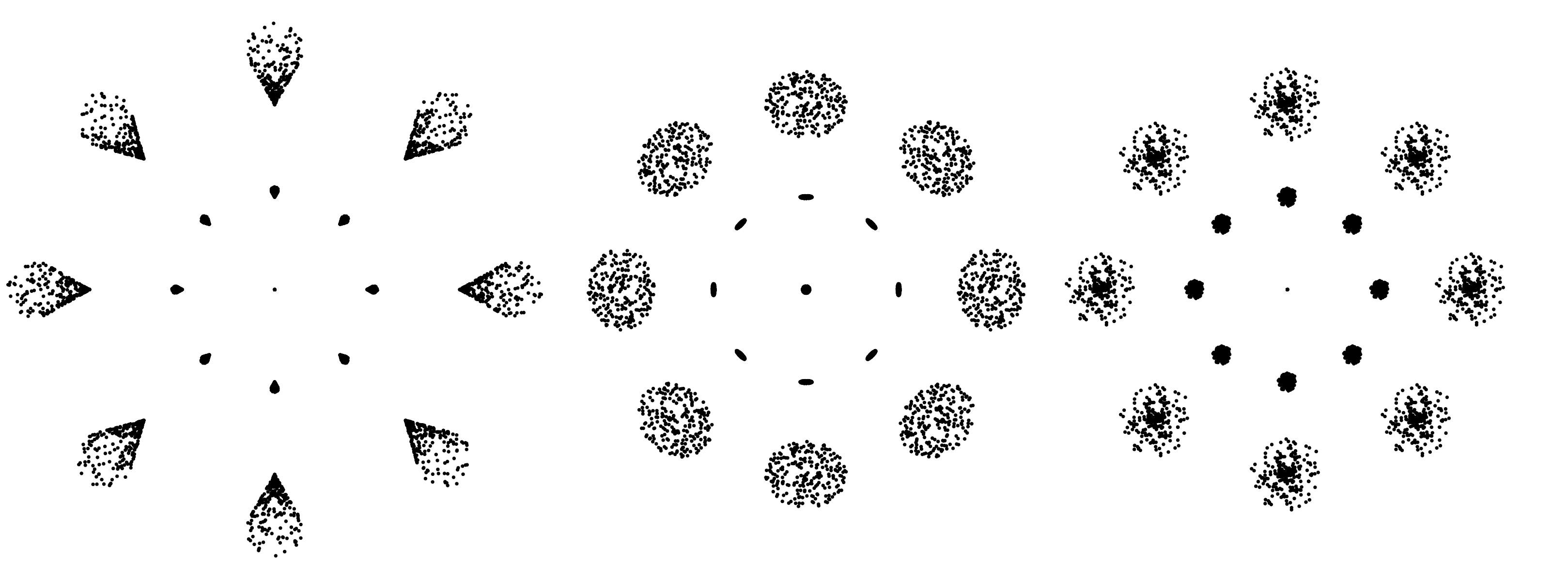}
   \end{tabular}
   \end{center}
   \caption[3Vsym] 
   {\label{fig:3Vsym} 
Symmetric 5th order aberration patterns: coma-I, astigmatism-I, and
spherical-I.  These differ from the patterns in Figure \ref{fig:3sym} only
in the dependence upon field angle and not in the character
of the point spread function.  Not shown is 5th order defocus.
}
   \end{figure}

\begin{figure}
   \begin{center}
   \begin{tabular}{c}
   \includegraphics[scale=0.52]{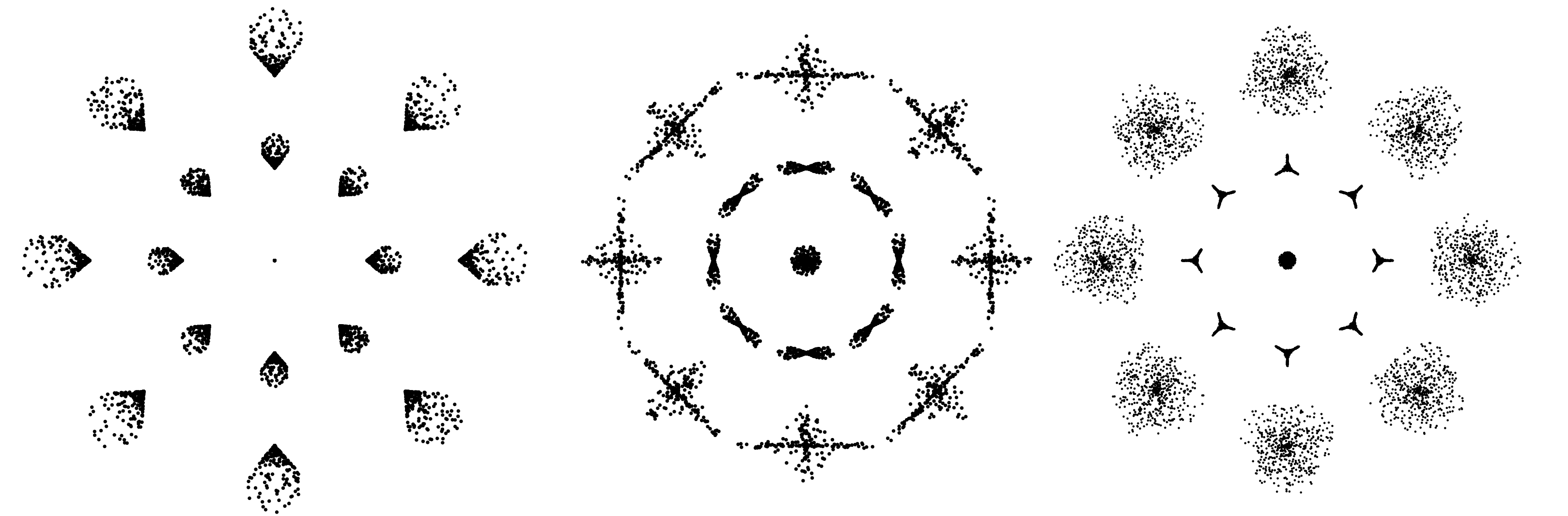}
   \end{tabular}
   \end{center}
   \caption[3IIsym] 
   {\label{fig:3IIsym} 
More symmetric 5th order aberration patterns: coma-II, astigmatism-II, and
trefoil.  Note that the character of the point spread function for
coma-II differs from that of the third order coma-I 
in figure \ref{fig:3sym}, and likewise that for astigmatism-II and astigmatism-I.
Not shown is spherical-II, which is constant across the field.
A constant spherical aberration  has been added to the astigmatism-II to
show the orientations of the images.  A constant aberration varying as
$\rho^3$ on the pupil has likewise been added to the trefoil.
}
   \end{figure}

In addition to the third order aberrations patterns of \S 2, despace
errors produce symmetric fifth order aberration patterns, six of
which are shown in Figures \ref{fig:3Vsym} and \ref{fig:3IIsym}.  
Not shown are spherical-II, which is
constant across the field, and fifth order defocus, for which
the dependence on field angle is the same as for fifth order astigmatism.
A more
complete discussion is presented by Schechter and Levinson\cite{Schechter2011}.

\subsection{Asymmetric Fifth Order Aberration Patterns}

\begin{figure}
   \begin{center}
   \begin{tabular}{c}
   \includegraphics[scale=0.52]{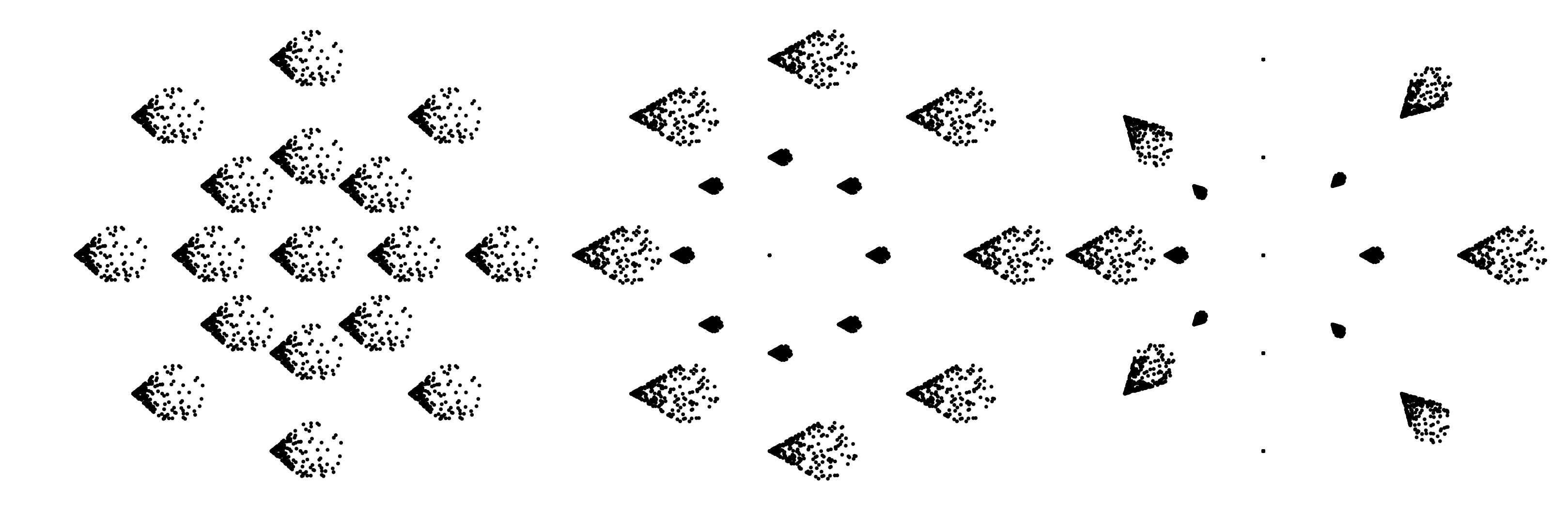}
   \end{tabular}
   \end{center}
   \caption[3coma] 
   {\label{fig:3coma} 
Asymmetric 5th order aberration patterns -- The pattern on the left
is for coma-II.  The pattern in the center and the one on the right are
both for coma-I.  A similar triplet of patterns might be shown for 
astigmatism-II and astigmatism-I.  There are many more patterns that
are not shown.
}
   \end{figure}

Tilt and decenter errors also produce asymmetric fifth order
aberration patterns.  Figure \ref{fig:3coma} shows {\it only} those associated with
the different types of coma.  Asymmetric fifth order coma-II is
constant across the field, as was the case for asymmetric third order
coma-I.  There are two distinct patterns for asymmetric fifth order
coma-I.  

Asymmetric fifth-order patterns for spherical, trefoil, defocus,
astigmatism-II and astigmatism-I can be found in Schechter and
Levinson.\cite{Schechter2011} As with asymmetric fifth order coma-I,
asymmetric fifth order astigmatism-I exhibits two distinct patterns,
each of which might be used to constrain two degrees of freedom.

\section{THE SUBSPACE OF BENIGN MISALIGNMENT}

Any one of the asymmetric fifth-order patterns presented in the previous
section might be used to give the two additional numbers that we have
seen are needed to achieve perfect alignment of a TMA.  But examination
of the {\it amplitudes} of these patterns shows them to be 
small, as shown by Schechter and Levinson\cite{Schechter2011} and in particular, 
in Tessieres'\cite{Tessieres2003}
treatment of an early version of the LSST.  These small amplitudes
make it difficult to achieve perfect alignment.

Conversely, these same small amplitudes imply that one need not strive
for perfect alignment.  If one eliminates the asymmetric third order
aberration patterns, the residual asymmetric fifth order patterns will
be small as long as one has not strayed too far from the telescope's
design.  

For a TMA there is a two dimensional space of misalignments that
produces only such fifth-order asymmetric patterns.  We refer to this
as the {\it subspace of benign misalignment} -- the subspace in
which asymmetric third order coma, astigmatism and curvature of field
are all zeroed.

\subsection{Ensuring Benign Misalignment}

To zero out asymmetric third order coma, astigmatism and curvature of field,
one must control six asymmetric degrees of freedom.  For our hypothetical
TMA this means controlling both the secondary {\it and} the tertiary.

But until now we have implicitly assumed a perfectly aligned focal
plane.  Asymmetric curvature of field manifests itself as a tilt of
the focal plane.  One might therefore, alternatively, control four
degrees of freedom on the secondary {\it and} the tilt of the focal plane.

\section{HOW MANY WAVEFRONT SENSORS AND HOW MANY ZERNIKE MODES?}

The most straightforward way to keep a TMA aligned would be to measure
symmetric first order defocus and third order spherical to control the
despace errors and to measure asymmetric coma, astigmatism and
curvature of field patterns to stay within the subspace of benign
misalignment.  The Zernike modes are then limited to the first eight.

Simultaneous measurement of first order defocus and asymmetric
curvature of field (which manifests itself as focal plane tilt)
requires {\it a minimum of three} wavefront sensors.  One might
alternatively use symmetric third order coma (or astigmatism) instead
of first order defocus, but the best way to control first order
defocus (which is large) is to measure it directly rather than infer
it indirectly.  

One would also want to measure at least one of the asymmetric fifth
order aberration patterns to be sure that one hasn't travelled too far
from the telescope design within the subspace of benign misalignment.
The choice of which pattern to measure will depend upon the expected
amplitudes of the patterns and the positions chosen for the wavefront
sensors.

Note, by way of example, that three wavefront sensors at the same
distance from the center of the field cannot discriminate between
asymmetric third-order coma and one of the two kinds of asymmetric
fifth-order coma-I, both of which are constant at fixed field angle.

Another consideration is the similarity of certain aberrations.
Coma-I and coma-II are quite similar, and will give correlated
measurements that in turn, will require high signal to noise to get
adequate discrimination.

\subsection{Mirror Deformations}

While we have concerned ourselves with mirror misalignments, wavefront
sensors are likely to be used both for maintaining alignments {\it and}
for maintaining mirror figures.  While we have shown that one 
need only measure low order Zernike modes to maintain alignment,
one may need higher order modes to maintain mirror 
figure\cite{Noethe1991, Schechter2003}.

It is fortunate that none of the misalignment patterns gives constant
astigmatism, because the softest bending modes for the primary mirror
will typically give such a signal\cite{Noethe2000}.  But the next softest
mode, at least in the case of a primary mirror with a central hole,
is likely to be cone-like.  Such a deformation, in combination with
defocus, looks much like spherical aberration\cite{Schechter2003}.
One must measure the wavefront to higher Zernike order to separate
these.


\bibliography{report}   
\bibliographystyle{spiebib}   

\end{document}